\documentstyle[aps,prl,floats,epsf]{revtex}
\begin{document}
\advance\textheight by 0.2in
\draft
\twocolumn[\hsize\textwidth\columnwidth\hsize\csname@twocolumnfalse%
\endcsname
\title{Weakly Pinned Bose Glass vs. Mott Insulator Phase in 
	Superconductors} 

\author{Carsten Wengel$^{1}$ and Uwe Claus T{\"a}uber$^{2}$}

\address{$^{1}$Department of Physics, University of California, 
	Santa Cruz, California 95064 \\ 
	$^{2}$Department of Physics --- Theoretical Physics,
	University of Oxford, 1 Keble Road, Oxford OX1 3NP, U.K.
	\\ and Linacre College, St. Cross Road, Oxford OX1 3JA, U.K.}

\date{\today}

\maketitle

\begin{abstract}
We study the properties of the Bose glass phase of localized flux
lines in irradiated superconductors near the matching field $B_{\Phi}$. 
Repulsive vortex interactions destroy the Mott insulator phase
predicted to occur at $B=B_\Phi$. 
For ratios of the penetration depth to average defect distance
$\lambda / d \leq 1$ remnants of the Mott insulator singularities
remain visible in the magnetization, the bulk modulus, and the
magnetization relaxation, as $B$ is varied near $B_{\Phi}$. 
For $\lambda \geq d$, the ensuing weakly pinned Bose glass is
characterized by a soft Coulomb gap in the distribution of pinning
energies.
\end{abstract}

]


For the application of high--$T_c$ (type II) superconductors in
external magnetic fields, an effective flux pinning mechanism is
essential in order to minimize the resistive losses through
Lorentz--force induced vortex motion.
Specifically, columnar defects, i.e., linear damage tracks in the
material caused by heavy--ion irradiation, have emerged as very
effective pinning centers \cite{blatter94}.
For such systems, a continuous vortex localization transition at
$T_{BG}$ from an entangled flux liquid to a disorder--dominated Bose
glass phase was predicted \cite{lyuksyutov92,nelson92}, and
subsequently found in experiment \cite{blatter94}.
In addition, it was suggested that when the vortex and defect
densities are equal, each flux line would be attached to one pin, 
leading to a Mott insulator phase within the Bose glass 
\cite{nelson92}.
Recent measurements of the magnetization relaxation rate at low
temperatures \cite{beauchamp95,nowak95}, and of the reversible
magnetization itself \cite{li96,beek96}, have been interpreted as
signatures of this Mott insulator, and inspired further theoretical
investigations \cite{radzihovsky95,reichhardt96,bulaevskii96}.
Yet so far, the influence of the repulsive vortex interactions, whose 
range is set by the London penetration depth $\lambda$, has not been
carefully studied.

\begin{figure}[t]
\epsfxsize=\columnwidth\epsfbox{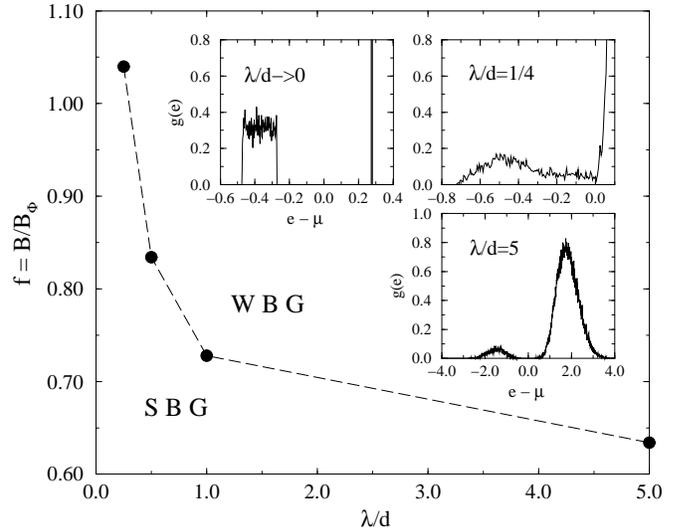}
\caption{SBG / WBG crossover line (see text) vs. $\lambda/d$.
	The insets show the pinning energy distribution $g(e)$
	vs. $e-\mu$ for $\lambda/d\to 0$ (left), $\lambda/d=1/4$
	(right top) and $\lambda/d=5$, all at $B_{\Phi}$.}
\label{occupation}
\end{figure}

The aim of this Letter is to explore numerically how the repulsive
vortex forces affect the predicted Mott insulator as the magnetic
field $B$ is varied near the matching field $B_{\Phi}$.
Using a {\it random} defect distribution with average defect distance
$d$, we extend earlier work which was limited to $B\ll B_{\Phi}$
\cite{taeuber95}.
Our findings are: 1) At $B_{\Phi}$ the Mott insulator, which is
characterized by a {\it hard} gap $\Delta$ in the distribution of
pinning energies $g(e)$ near the chemical potential $\mu$, exists only
for extremely short--range interactions $\lambda/d \to 0$ 
(Fig.~\ref{occupation}, left inset); with increasing $\lambda<d$ the 
gap quickly fills and, therefore, the Mott insulator is destroyed
(Fig.~\ref{occupation}, right top inset). 
For long--range interactions $\lambda>d$ a {\it soft} Coulomb gap
\cite{shklovskii84} described by $g(e)\propto \vert e - \mu
\vert^{s_{\rm eff}}$ emerges 
with an effective gap exponent $s_{\rm eff}$ 
(Fig.~\ref{occupation}, bottom inset).
2) By measuring the reversible magnetization and the effective IV
exponent, which is related to the magnetization relaxation rate, we
are able to explain experimental results
\cite{beauchamp95,nowak95,li96,beek96} as {\it remnants} of the Mott
insulator for $\lambda<d$, but {\it not} as true signatures of this
distinct thermodynamic phase itself.
3) We can identify a crossover line (Fig.~\ref{occupation}),
discriminating between the strong Bose glass (SBG, with all vortices 
localized by defects) and the weak Bose glass (WBG, with vortex bundle
pinning) \cite{nelson92,radzihovsky95}.

The theoretical description of the Bose glass is based on the
following free energy of $N_V$ flux lines, described by their 2d
trajectories ${\bf r}_i(z)$ as they traverse a sample of thickness $L$
in a magnetic field ${\bf B} \parallel {\bf \hat z}$
\cite{lyuksyutov92,nelson92},
\begin{eqnarray} \label{free_energy}
{\cal F} & = & \int_0^L \! dz \sum_{i=1}^{N_V} 
\left \{ \frac{\tilde{\epsilon}_1}{2} \left( \frac{d{\bf r}_i (z)}
{dz}\right)^{\! 2} + \frac{1}{2} \sum_{j\neq i}^{N_V} V[r_{ij}(z)]
\right. \nonumber \\ & &\qquad \qquad \qquad + \left. 
\sum_{k=1}^{N_D} V_D[{\bf r}_i (z) - {\bf R}_k (z)] \right \}\ .
\end{eqnarray}
This consists first of an elastic line tension term, with tilt modulus 
$\tilde{\epsilon}_1$. 
The second term denotes the interaction energy of all vortex pairs
(local in $z$), where $r_{ij}=\vert {\bf r}_i - {\bf r}_j \vert$ and 
$V(r)=2 \epsilon_0 K_0 (r/\lambda)$ is the screened repulsive vortex
potential, with the modified Bessel function $K_0(x)\propto -\log(x)$
as $x\to 0$, and $K_0(x)\propto x^{-1/2}\exp(-x)$ for $x\to\infty$. 
The energy scale is set by $\epsilon_0=(\phi_0/4\pi \lambda)^2$.
The last term describes $N_D$ columnar pins ($\parallel {\bf B}$),
modeled by $z$--independent square well potentials $V_D$ with average
spacing $d$, centered on randomly distributed positions 
$\{{\bf R}_k\}$.  
The typical defect radius is $c_k \approx 50 \AA$, and related to the
pinning strengths $U_k$ via $U_k \approx (\epsilon_0/2) 
\log[1 + (c_k/\sqrt2 \xi)^2]$ \cite{nelson92}.
The ion beam dispersion induces a distribution of the $U_k$ with
width $w=\sqrt{\langle \delta U_k^2 \rangle}$.

The mathematical analysis of Eq.~(\ref{free_energy}) exploits a
mapping of the statistical mechanics of this free energy of directed
lines to the quantum mechanics of two--dimensional bosons subject to
point disorder \cite{lyuksyutov92,nelson92}.
Here, we focus on the ground states properties of
Eq.~(\ref{free_energy}) for $B\sim B_{\Phi}$ at zero temperature,
which is a fair approximation for the Bose glass phase for 
$T<T_1\approx 0.6 \ldots 0.8 \, T_{\rm BG}$.  
In this regime, thermal wandering of vortices can be neglected 
\cite{nelson92,taeuber95,krusin96,baumann96}, and the vortices will
essentially be straight which allows us to ignore the elastic energy.
This leaves us with a static two--dimensional problem of $N_V$
interacting ``particles'' and $N_D$ defects.
For computational reasons we represent our problem on an underlying
triangular grid with $N$ ``lattice'' sites and therefore $N-N_D$
interdefect sites (``interstitials'').
The effective Hamiltonian then reads
\begin{equation} \label{eff_hamiltonian}
{\cal H_{\rm eff}}= \frac{1}{2} \sum_{i\neq j}^N n_i n_j V(r_{ij}) +
\sum_{k=1}^{N_D} n_k t_k \ ,
\end{equation}
where $\{n_i=0,1\}$ represent the site occupation numbers.
If we take $\xi=10\AA$ and $c_0=50\AA$, we obtain for the ``bare''
defect pinning energies $t_k=-\langle U_k \rangle + w_k$, with 
$\langle U_k \rangle= 0.65$ and width $w=0.1$ in units of
$2\epsilon_0$. 
For simplicity we assume a flat distribution of pinning energies
around its mean, $P(w_k)=\Theta (w-\vert w_k\vert)/2w$, where $\Theta$
denotes the step function.
The simulations are performed by randomly distributing $N_V$ vortices
and $N_D$ defects on the triangular grid and minimizing the total
energy with respect to single--particle transfers, thereby obtaining
pseudo--groundstates for the Hamiltonian (\ref{eff_hamiltonian})
\cite{footnote0}.
Our simulations are mostly carried out with $N=1600$, $N_D=N/16=100$,
using periodic boundary conditions. We study physical quantities
as functions of the interaction range $\lambda/d=1/4,1/2,1,5$, and the
filling fraction $f=N_V/N_D=B/B_{\Phi}$, in the interval $0.2\leq f\leq 3$.
All distances are given in units of the lattice constant $a$ of the
triangular grid.
The lower limit for the interaction range is determined by 
$\lambda=a$, since even smaller values effectively correspond to
$\lambda\to 0$.
We typically take an average over $50$ different realizations of the
disorder.

We have compared our findings for $f \leq 0.6$ with those obtained in
Ref.~\cite{taeuber95}, where it was assumed that all vortices remain
bound to defects, irrespective of the value of $\lambda$, and
therefore continuously spaced random positions could be used.
Indeed, we were able to reproduce the previous results for filling
fractions $f=0.2$ and $f=0.4$ quantitatively. 
We have also tried to render the artificial grid finer by keeping
$N_D$ constant and increase $N$ from $1600$ to $3600$, the largest
sample we could study, and did not detect any lattice dependence for
both grid sizes. 
Hence, we believe that our results should be largely insensitive to the
underlying lattice representation, and provide a fair approximation to
a more realistic continuum description.
Also, we could not find any significant finite--size effects when
keeping $N=3600$ and $\lambda/d$ fixed and
increasing $N_D$ from $36$ to $100$ and $225$.

A natural first question to ask is how many flux lines are depinned as
a result of the vortex interactions, depending on the values of $f$
and $\lambda/d$. 
Within the Bose glass one may discriminate between the SBG and the
WBG, where the latter is characterized by a markedly reduced
localization temperature $T_{\rm BG}$ and critical current $J_c$ 
\cite{nelson92}.
For $\lambda/d \ll 1$, the crossover between these regimes
is expected to occur for $B\sim B_\Phi$\cite{radzihovsky95}.
Once interactions become strong, the WBG will appear well
below $B_\Phi$.
Fig.~\ref{occupation} shows the filling fraction $f_{\rm occ}$ at
which 10 \% of the vortices are depinned, as a function of $\lambda/d$,
which we tentatively take as a criterion to define a crossover line
between the SBG and WBG.
One observes a rather strong dependence of this line on $\lambda/d$,
which is shifted well below $B_\Phi$ as soon as $\lambda \approx d$.
Only for $\lambda/d\le 1/4$ does the line remain above $B_{\Phi}$; yet
obviously this is a prerequisite for the Mott insulator phase to
appear.

In order to see if the Mott insulator persists for a finite
interaction range {\it and} a random distribution of pinning sites, we
compute the single--particle density of states $g(e)$, i.e. the
distribution of (interacting !) pinning energies, where 
$e_i = \sum_{j\not=i} n_j V(r_{ij}) + t_i$.
For long--range interactions ($\lambda \gg d$) one expects a
soft ``Coulomb'' gap in $g(e)$ at the chemical potential $e=\mu$,
separating occupied ($e<\mu$) from empty ($e>\mu$) states; near $\mu$,
$g(e)$ should vanish according to $g(e)\propto \vert e-\mu
\vert^{s_{\rm eff}}$ \cite{shklovskii84,taeuber95}.
In the limit $\lambda/d\to 0$, on the other hand, the Mott insulator
phase at $f=1$ is characterized by the appearance of a hard gap
separating the occupied defect states at $e=-\langle U_k\rangle \pm w$
from the unoccupied states at $e=0$ (Fig.~\ref{occupation}, left
inset).
The right insets in Fig.~(\ref{occupation}) show $g(e)$ at $f=1$ for 
$\lambda/d=1/4$ (top) and $\lambda/d=5$ (bottom). 
The bottom figure exhibits a wide Coulomb gap, which we also find
(though narrower) for $\lambda/d=1$.

The top inset, however, only shows a fairly flat density of states
for the occupied sites, which is somewhat depleted for $e\approx \mu$,
and rises sharply for $e\ge\mu$.
The fact that all states below $\mu$ are {\it continuously} filled
implies that even rather short--range interactions suffice to
destabilize a distinct thermodynamic Mott insulator phase 
\cite{reichhardt96}.
We have also looked at $\lambda/d=1/8$ by increasing $N$ to $3600$,
using only $N_D=56$, whence $d=\sqrt{N/N_D}\approx 8$, and even here
we still found some states in the vicinity of $\mu$. 
We are therefore led to conclude that for the true Mott insulator to
appear in this system with a {\it random} spatial distribution of
pinning sites, a necessary condition is $\lambda\ll d$.  
We speculate that for a more regular array of columnar defects
\cite{baert95}, the Mott insulator at $T=0$ may persist to
considerably larger interaction ranges \cite{reichhardt96}.

\begin{figure}[t]
\epsfxsize=\columnwidth\epsfbox{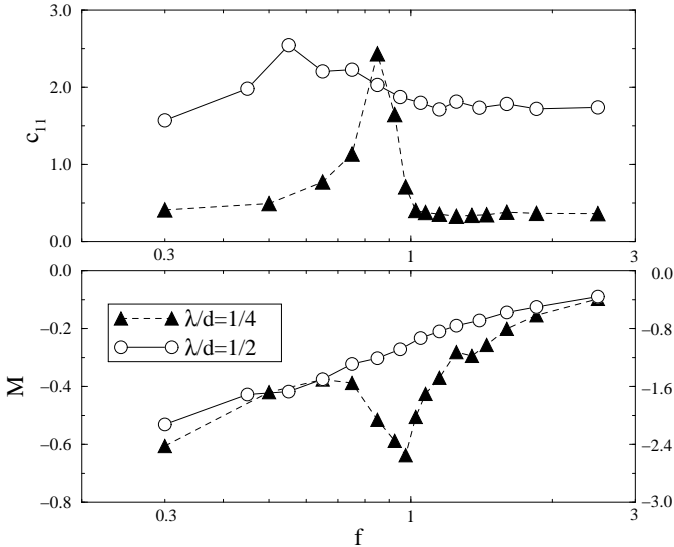}
\caption{Log--linear plots of the bulk modulus $c_{11}$ (top) and the
	 magnetization M (bottom) vs. $f$ for $\lambda/d=1/4$ (left
	 scale in the bottom plot) and $\lambda/d=1/2$ (right scale).} 
\label{bulk_mag_fig}
\end{figure}

Next, we have determined the chemical potential $\mu$ as a function of
$f$, i.e. the $H(B)$ curve, defining $\mu=(e_{\rm max}+e_{\rm min})/2$
with $e_{\rm max}$ the maximum site energy of the occupied states, and
$e_{\rm min}$ the minimum of the unoccupied site energies, and then
averaging over disorder \cite{shklovskii84}.
From this we obtain the bulk modulus of the system via 
$c_{11}= \partial \mu / \partial f = N_D \partial \mu / \partial N_V$.
A divergence of $c_{11}$ at $B_{\Phi}$ would indicate the appearance
of the Mott insulator, which is incompressible (adding another vortex
costs an energy $\Delta$) and in this respect similar to the Meissner
state.
Correspondingly, its signature should be a sharp jump in $\mu$ at
$f=1$ [for $\lambda\to 0$ the height of this jump is given by the hard
gap $\Delta = \langle U_k \rangle$ in g(e)]. 
Fig.~\ref{bulk_mag_fig} (top) shows that while the divergence is
smeared out already for short--range interactions ($\lambda/d=1/4$),
yet a quite pronounced peak in $c_{11}$ remains, shifted downwards to
$f\approx 0.85$.
For larger interaction range ($\lambda/d=1/2$) the jump of $\mu$ is
displaced to even smaller $f (\approx 0.55$) and less marked; for
$\lambda\ge d$, the bulk modulus becomes essentially constant over the
entire range of $f$.
In both cases depicted here, the location of the peak in $c_{11}$
coincides with that value of $f$ at which a sizeable number of vortices
actually leave the defect sites. 
The fact that the average bulk modulus is enhanced for larger
$\lambda$ is simply an energetic effect, since a vortex entering the
system has to overcome a higher energy barrier due to the
interactions.
The behavior of $c_{11}$ confirms again that no true Mott insulator
exists even for relatively short--range interactions
$\lambda/d=1/4$ or $1/2$.
We do observe, though, a distinctive "lock--in" structure for
$\lambda/d < 1$, which completely disappears for $\lambda/d\ge 1$. 

In order to make further contact with experiments, we measured the
total energy $G$, which yields the reversible magnetization via the 
thermodynamic relation 
$M=-\partial G/\partial B\propto -N_V^{-1}\partial G/\partial f$.
Fig.~\ref{bulk_mag_fig} (bottom) depicts $M$ as a function of $f$
on a log--linear plot for $\lambda/d=1/4$ and $\lambda/d=1/2$. 
The data for $\lambda/d=1/4$ show a pronounced minimum in $M$ at
$f\approx 1$, embedded in a slow logarithmic growth as $f$ increases. 
The second plot for $\lambda/d=1/2$ hardly displays any structure
aside from a shallow dip near $f\approx 0.6$. 
This feature is completely absent for $\lambda/d\ge 1$, where only the
$\log(\sqrt{f})$ increase remains, resembling the magnetization curve
of an unirradiated superconductor \cite{li96,beek96,bulaevskii96}.
The observed behavior of $M$ at $\lambda/d=1/4$ qualitatively agrees
very well with recent measurements performed by van der Beek 
{\it et al.} in an irradiated BSCCO crystal \cite{beek96}, who find 
(at $T\approx T_1$) a pronounced dip in $M$ centered near 
$B=B_{\Phi}$. The disappearance of this minimum in the experiments as
$T\to T_{\rm BG}^-$ may be at least partially due to the increase of
the London penetration depth $\lambda(T)$, in addition to the
entropic renormalizations studied in Ref.~\cite{bulaevskii96}.
Our simulations cannot explain, however, the magnetization minima
found at $f > 1$ in BSCCO tapes \cite{li96,footnote2}.

We would like to point out that the maximum in $c_{11}$ occurs at {\it
lower} values of $f$ than the minimum in $M$ (compare, e.g., our data
for $\lambda/d=1/4$), which can be understood as follows. 
Starting from the relation $M \propto B - H$, the bulk modulus may be
written as $c_{11} \propto \partial H/\partial B\propto {\rm const.}-
\partial M/ \partial B = {\rm const.}+\partial^2 G/\partial B^2$; thus
the maximum of $c_{11}$ occurs at the location of the steepest
negative slope in $M(B)$, which has to be at a smaller $B$ than the
minimum in $M$. 
Only in the ``true'' Mott insulator phase will the singularities both
in $c_{11}$ and $M$ coincide at $B=B_\Phi$.

We now briefly return to the case of long--range interactions 
$\lambda/d>1$, where a soft Coulomb gap appears in the distribution of
pinning energies (r.h.s. inset in Fig.~\ref{occupation}).
Notice that the energetically favorable sites include both the defect
and high--symmetry interstitial positions; thus, the original spatial
randomness is in effect smoothened, and the gap exponent $s_{\rm eff}$
{\it increased} as compared to a situation where no interstitials 
are allowed \cite{taeuber95}.
At large $f$, a substantial fraction of the vortices will leave the
pin positions, and as a consequence, Bose glass behavior can only be
seen at low temperatures \cite{baumann96}.
Yet, then the pinning via the repulsive forces exerted by neighbors
that are still attached to columnar defects is very effective; we find
that the potential minima are roughly equally deep and wide for all
vortices in a given pseudo--ground state configuration, irrespective
of the flux line occupying a defect or an interstitial site.
This gives us confidence that the single--vortex density of states may
indeed be used to infer low--current transport properties in the
variable--range hopping regime \cite{nelson92,taeuber95}.

\begin{figure}[t]
\epsfxsize=\columnwidth\epsfbox{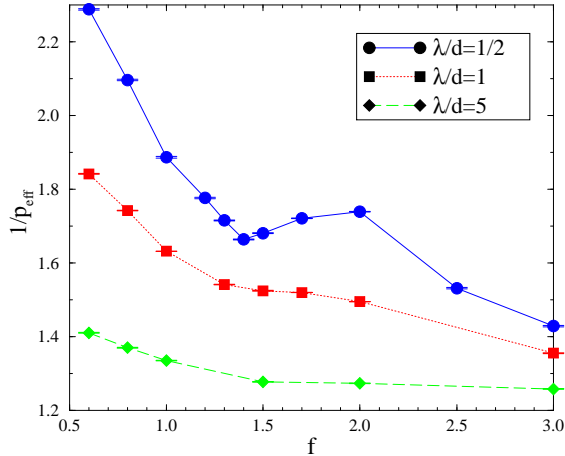}
\caption{Transport exponent $1/p_{\rm eff}$ vs. $f$ for
	$\lambda/d=1/2,1,5$.} 
\label{peff-pl}
\end{figure}

In the localized Bose glass phase, vortex transport at low currents 
$J \ll J_c$ is expected to occur via variable--range hopping
\cite{nelson92}, in analogy with doped semiconductors
\cite{shklovskii84}.
In the spirit of the thermally--assisted flux flow model, this leads
to a highly nonlinear IV characteristics 
${\cal E} = \rho_0 \, J \, \exp [ - U_{\rm B}(J) / k_{\rm B} T ]$,
with effective (free) energy barriers that diverge according to 
$U_{\rm B}(J) = U_0 \, (J_0 / J)^{p_{\rm eff}}$ as $J \to 0$. 
For short--range interactions, $p_{\rm eff}$ should be given by the 2d
Mott variable--range hopping exponent $p_0 = 1/3$ \cite{nelson92}. 
Yet, once the vortex interactions become long--range, the emergence of a 
Coulomb gap in the distribution of pinning energies near $\mu$ leads
to a considerable enhancement of flux pinning with effective exponents
up to $p_{\rm eff} \approx 0.7$, if $f=0.1$ \cite{taeuber95}.
Such values for $p_{\rm eff}$ in the range between $1/3$ and $1$ (as
suggestive of the vortex half--loop excitations dominating for
intermediate currents \cite{nelson92,taeuber95}) were found in recent
magnetization relaxation experiments extending up to $B_{\Phi}$
\cite{baumann96}.

We have calculated $p_{\rm eff}$ from the IV characteristics obtained
by integrating over $g(e)$ for $e>\mu$ (see Ref.~\cite{taeuber95}).
In Fig.~\ref{peff-pl} we plot $p_{\rm eff}^{-1}$ as a function of $f$
for various values of $\lambda/d$.
Remarkably, $p_{\rm eff}$ {\it increases} with growing vortex density
for $f<1$, as opposed to the earlier simulations where no interstitial
positions were allowed and to the contrary a ${\it decrease}$ with
$f$ was found, owing to the fact that the system had to accommodate
with the underlying randomness \cite{taeuber95}.
Now, however, with favorable interstitial sites being available, the
disorder effects are screened by the interactions as the vortex
density increases.
This leads to stronger correlations, and in fact more mean--field like
behavior with larger values of $p_{\rm eff}$ as is apparently
seen in experiment \cite{baumann96}.
For $\lambda/d=1/2$, we find a marked minimum in $p_{\rm eff}^{-1}$ at 
$f \approx 1.4$ (see Fig.~\ref{peff-pl}) obviously due to a delicate
interplay of disorder and correlation effects.  
Assuming that relaxation times are determined by the variable--range
hopping energy barriers (vortex superkinks), 
$\log (t/t_0) \approx U_{\rm B}(J)/k_B T$ \cite{blatter94}, one finds
for the magnetization relaxation rate 
$S=- dM/d\log t=-d\log J/d\log t = k_{\rm B}T/p_{\rm eff} U(J)$;
therefore $S(B) \propto 1 / p_{\rm eff}(f)$.
Hence, the minima in $S(B)$ near $B \approx 1.4 B_\Phi$ detected by
low--temperature relaxation experiments on YBCO \cite{beauchamp95},
and on a Tl--compound \cite{nowak95}, might be explained by the
maximum in $p_{\rm eff}$ we find with our simulations.
At elevated temperatures, this structure should disappear quickly in
this WBG regime.

In summary, we have investigated the Bose glass phase of flux lines
localized by columnar defects near the matching field, taking the
vortex interactions properly into account.
For a random spatial distribution of pinning centers, the proposed
Mott insulator phase is destabilized by the repulsive forces between
the flux lines, but for a moderate interaction range $\lambda/d < 1$
interesting ``lock--in'' effects remain visible in the reversible
magnetization, the bulk modulus, and the magnetization relaxation,
which at least qualitatively explains a number of recent experiments.
For larger $\lambda/d > 1$, the distribution of pinning energies
displays a wide Coulomb gap, and sites available for variable--range
hopping processes include both defect as well as high--symmetry
interstitial positions.

We benefitted from discussions with M. Baumann, C.J. van der Beek,
J.T. Chalker, J. K\"otzler, L. Radzihovsky, and A.P. Young.
C.W. acknowledges support from a NSF Grant DMR 94--11964 and
U.C.T. through a European Commission TMR Grant ERB FMBI-CT96-1189.


\begin{thebibliography}{99}
\bibitem{footnote0}
  Details of this minimization procedure are described in
  Refs.~\cite{taeuber95,shklovskii84}, and further references therein.
\bibitem{footnote2}
  One possibility might be unspecified correlations in the spatial
  defect distribution in the sample.
\end{thebibliography}
\end{document}